\pdfoutput=1
\documentclass[11pt]{article}

\usepackage[]{ACL2023}

\usepackage{times}
\usepackage{latexsym}
\usepackage{graphicx}% http://ctan.org/pkg/graphicx

\usepackage[T1]{fontenc}
\usepackage[utf8]{inputenc}

\usepackage{microtype}

\usepackage{inconsolata}

\usepackage{graphicx}
\usepackage{tcolorbox}
\usepackage{amsmath}

\title{GPT-4 is judged more human than humans in displaced and inverted Turing tests}
\author{Ishika Rathi \and Sydney Taylor \and Benjamin K. Bergen \and Cameron R. Jones \\
Department of Cognitive Science, UC San Diego \\
        9500 Gilman Dr, San Diego, CA, USA\\
        \texttt{\{irathi,cameron\}@ucsd.edu}}

\begin{document}
\maketitle

\begin{abstract}
Everyday AI detection requires differentiating between people and AI in informal, online conversations. In many cases, people will not interact directly with AI systems but instead read conversations between AI systems and other people. We measured how well people and large language models can discriminate using two modified versions of the Turing test: \textit{inverted} and \textit{displaced}. GPT-3.5, GPT-4, and displaced human adjudicators judged whether an agent was human or AI on the basis of a Turing test transcript. We found that both AI and displaced human judges were less accurate than interactive interrogators, with below chance accuracy overall. Moreover, all three judged the best-performing GPT-4 witness to be human more often than human witnesses. This suggests that both humans and current LLMs struggle to distinguish between the two when they are not actively interrogating the person, underscoring an urgent need for more accurate tools to detect AI in conversations. 
\end{abstract}

\section{Introduction}

In 1950, Alan Turing devised the \textit{imitation game} as a test to indirectly investigate the question, "Can machines think?" In a classic Turing test, a human interrogator engages in a text-only conversation with two witnesses: one human and one machine. If the interrogator is unable to accurately differentiate between the human and the computer, the computer passes the test and can be considered intelligent. Since Turing's original paper, the Turing test has sparked an intense debate that has been pivotal in constructing modern understandings and conceptions of intelligence, shaping the fields of computer science, cognitive science, artificial intelligence, robotics, philosophy, psychology, and sociology \cite[p. 116]{french2000turing}.

Beyond its controversial role as a test of intelligence, the Turing test also serves as a measure of whether humans can detect AI in conversational settings, or whether AI models can successfully deceive human interlocutors into thinking that they are human. 
Recent empirical work has found that interrogators could not reliably determine whether a GPT-4-based agent was human or AI in a Turing test \cite{jones_people_2024}. 
Models that can successfully impersonate people bring attendant risks. This motivates conducting variations of the Turing test in more ecologically-valid settings to determine how effective people are in discriminating between humans and AIs in realistic scenarios.
An ordinary Turing test provides the interrogator with a key advantage not always present in passive consumption of AI-generated text: they can adapt their questions to adversarially test the witness in real time.
Here, we ask how well human and AI judges perform without this advantage, when they only have access to a transcript of a Turing test interview conducted by a separate participant.

\begin{figure}[]
  \includegraphics[width=\linewidth]{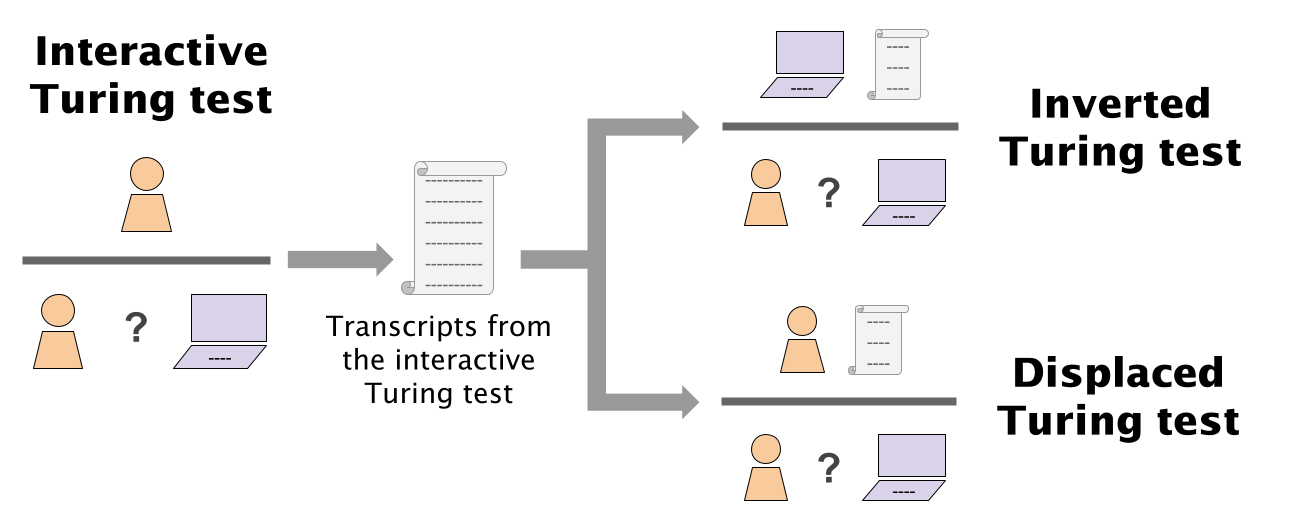}
  \caption{A summary of our experimental design. Transcripts were sampled from an interactive Turing test, where a human judge interrogates a witness to determine if they are human or AI. In an \textit{inverted} Turing test, we present transcripts to AI models, who judge whether the same witnesses are human or AI. In a \textit{displaced} Turing test, a separate group of human participants read the same transcripts and make this judgement.}
  \label{fig:experiment_setup}
\end{figure}

\subsection{Interactive Turing Test}
A classic Turing test involves a human evaluator interactively interrogating a witness to determine whether they are human or AI. 
Although the Turing test was originally proposed as a test of intelligence, there have been a wide variety of objections to its validity or sufficiency in this guise \cite{hayes1995turing, marcus_am_2017, french2000turing, oppy_turing_2003}. Independent of its validity as a measure of intelligence, the Turing test provides a powerful test for assessing similarities between human- and AI-writing and a useful premise for studying AI deception \cite{park_ai_2023}. 

Several attempts have been made to pass the Turing test, including the Loebner Prize---a competition that ran from 1990-2020 without any system passing \cite{shieber1994lessons};  ``Human or Not'', a large-scale social Turing test experiment that found an interrogator accuracy rate of 60\% \cite{jannai_human_2023}; and a 2024 study reporting the first system to have a pass rate statistically indistinguishable from chance (54\%) but still short of the human threshold (67\%) \cite{jones_people_2024}. Several variations of the test exist, with each informing dimensions of theory and practice. 

\subsection{Inverted Turing Test}
The first of these variations is the inverted, or reverse Turing test, which places an AI system in the role of the interrogator. \citet{watt_naive_1996} proposed the inverted test as a measure of \textit{naive psychology}, the innate tendency to recognize intelligence similar to our own and attribute it to other minds. It is passed if an AI system is "unable to distinguish between two humans, or between a human and a machine that can pass the normal Turing test, but which can discriminate between a human and a machine that can be told apart by a normal Turing test with a human observer,” \cite[p. 8]{watt_naive_1996}. Watt argued that by placing AI in the observer role and comparing its accuracy for different witnesses with human accuracy, the AI would reveal whether it has naive psychology comparable to humans.

As AI systems create larger proportions of online content \citep{fagni2021tweepfake}, and interact with others as social agents \citep{sumers2023cognitive}, the inverted Turing test takes on new real-world relevance. AI systems are already being used to discriminate between humans and bots online, for example, through the widespread implementation of CAPTCHA (Completely Automated Public Turing test to tell Computers and Humans Apart), reCAPTCHA, or invisible CAPTCHA \cite{yamamura_2013, pal2020deceiving}. The growing role of AI agents in online interactions raises questions around how well these systems will be able to discriminate between human and AI-generated content, and what kinds of criteria they might use to do so.

\subsection{Displaced Turing Test}
Several studies have assessed how well humans are able to recognize displaced AI-generated content in different domains including higher education \citep{perkins_detection_2024}, news \citep{moravec_human_2024}, online content \citep{cooke_as_2024}, images \citep{somoray_providing_2023}, and academic articles \citep{gao_comparing_2023, casal_can_2023}. 
Though these discrimination tasks bear similarities to the Turing test, there remain important differences. First, these tasks can only be considered a ``static'' version of the test, as the judgement is based on pre-existing and unchanging content generated fully by a human or an AI. Second, while an interactive interrogator in a traditional Turing test can ask dynamic, flexible, and adversarial questions, the judge in a static Turing test can only consider what an agent happened to say, and cannot interact to pursue the most fruitful lines of questioning.
Though static tests are therefore more limited in scope as tests of model abilities, they are likely to be parallels of a much more frequent occurrence in the real world, as many interactions are read by a larger audience than the addressee.
Here, we introduce a novel kind of static Turing test called a \textit{displaced} Turing test, wherein a human judge reads a transcript of an interactive Turing test that was previously conducted by a different human interrogator. The new human judge is ``displaced'' in that they are not present to interact with the witness.

\subsection{Statistical AI-detection methods}

There exist a variety of statistical approaches to detecting AI-generated content. These are largely based on the principle that LLMs generate content by sampling from a probability distribution over words which may leave particular probabilistic signatures, such as LLM generations being statistically more probable than human-generated ones \citep{ippolito-etal-2020-automatic, solaiman2019release, gehrmann-etal-2019-gltr}. \citet{mitchell2023detectgpt} developed a related metric, \textit{curvature}, which measures the local optimality of a piece of text with respect to small perturbations generated using a masked language model; LLM-generated content is likely more probable than nearby perturbations.
\citet{mireshghallah2024smaller} found that smaller LLMs tend to be better detector models, with a 125m parameter OPT model performing best at detecting AI-generated content overall, and achieving 90\% accuracy on GPT-4 specifically.

Various commercial tools have been developed on the basis of these methods and other computational approaches to classifying text. Studies have examined the effectiveness of these approaches in different settings with mixed results: while accuracy is high in some domains, models are unreliable and are likely to issue false positives \citep{elkhatat_evaluating_2023, bellini_between_2024, perkins_detection_2024}. Studies have also considered LLMs as detection tools in a question-answering setting. \citet{bhattacharjee_fighting_2024} assessed GPT-4’s and ChatGPT’s ability to discriminate between human-written and AI-generated content, and reported higher detection accuracy with human writing than with AI writing.

\subsection{Present Study}
These variations of the Turing test present the opportunity to ask new questions with real-world relevance.
In practice, human users often do not directly interact with bots but instead read their conversations with other human or AI users. As such, displaced human judges are important parallels for everyday AI detection. 
Additionally, the inverted Turing test provides a tool to investigate LLMs' latent model of human minds, and of their practical effectiveness in detecting AI-generated content online. If AI is used for detection, this becomes increasingly important. Mitigating technological deception will require ensuring that AI technologies for detection stay ahead of technologies for deception \citep{park_ai_2023}. 

We conducted Turing test variations by sampling transcripts from a previously-run interactive Turing test \cite{jones_does_2023}, and presenting them to displaced AI and human judges (Figure \ref{fig:experiment_setup}). We used this paradigm to address several questions: Can humans reliably discriminate humans from AIs from merely observing conversations? Can LLMs serve as AI detectors, not only for static pieces of writing (essays, articles, \& paragraphs) but also dynamic conversations? Does displacing the Turing test increase or decrease accuracy? Does an inverted Turing reveal naive psychology in artificial systems? And what methods work best for AI detection in real-world conversational spaces?

\section{Study 1: Inverted Turing Test}
We first place GPT-3.5 and GPT-4 in the role of the evaluator in an inverted Turing test.
Because the models are not actively interrogating the witness, we refer to this more passive role as the \textit{adjudicator}.
We presented both models with transcripts of previously-run Turing tests \citep{jones_does_2023} and elicited judgements about whether the witness was a human or an AI. We pre-registered 5 hypotheses on \href{https://osf.io/g6h4d/}{OSF}.
First, we asked whether GPT-4 accuracy would be significantly greater or less than human accuracy.
Second, in the original study, interactive human interrogators were no better than chance at determining whether one GPT-4 witness was human or AI. We asked whether GPT-4 would also fail to accurately identify this witness.
Third, in order to test whether GPT-4 could distinguish between real humans and well-designed systems, we asked whether the model would judge real humans to be human more often than it judged the best GPT-4 witness to be human.
Fourth, in order to understand the role that the length of the conversations plays in their identifiability, we asked whether GPT-4 accuracy as an adjudicator would be higher for longer transcripts.
Finally, to investigate the role of LLM model quality in judging transcripts, we asked whether GPT-4 would have higher accuracy as an adjudicator than GPT-3.5.
\begin{figure}[ht]
\begin{center}
\includegraphics[width=0.99\linewidth]{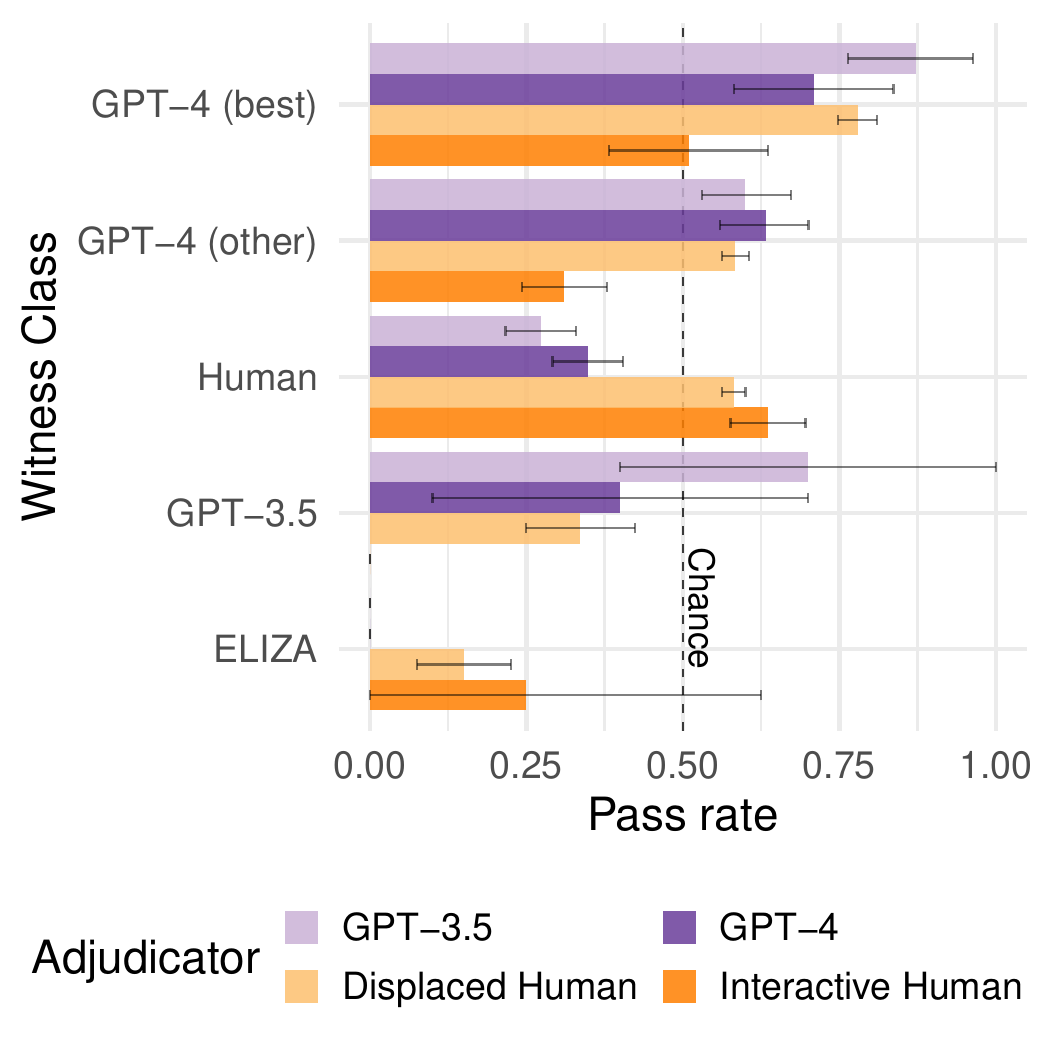}
\end{center}
\caption{
Mean pass rates (the proportion of time witnesses were judged to be human) by witness and adjudicator types. AI adjudicators (GPT-3.5 and GPT-4) judged GPT-4 witnesses to be human more often than they did real human witnesses. For displaced human adjudicators this was only true for the best GPT-4 witness. ELIZA's pass rate was low across all adjudicators.
} 
\label{fig:pass_rates}
\end{figure}

\subsection{Methods}
We sampled 500 transcripts from \citet{jones_does_2023} who conducted more than 6,000 Turing tests with interactive human adjudicators. Our sample was constrained to ensure i) an equal number of human and AI witnesses, ii) the same proportion of games from the best-performing GPT-4 witness as the full dataset, and iii) an equal distribution of human and AI verdicts for the best-performing model. We presented the transcripts to two leading AI models (GPT-3.5 and GPT-4), followed by a prompt that instructed models to make a judgement on whether the witness was a human or AI, to indicate a confidence level in the verdict, and to provide reasoning for judgements (Appendix \ref{appendix:instructions_ai}). 

AI witnesses in \citet{jones_does_2023} were based on 3 models: GPT-3.5, GPT-4 \citep{achiam2023gpt}, and ELIZA \citep{weizenbaum1966eliza}. Handwritten prompts instructed LLM-based systems on how to respond to interactive interrogators, with each prompt creating a unique AI witness. ELIZA was included as a performance baseline and a manipulation check: if participants successfully determine that ELIZA is AI---but not other AI witnesses---pass rates for AI witnesses are unlikely to be explained by a simple proclivity to anthropomorphism. 
We used the OpenAI API to access GPT-3.5 and GPT-4. Our analyses used the \textit{lmertest} package in R to fit linear mixed effects models to determine whether effects were statistically significant.

\subsection{Results}

We compared the accuracy of AI adjudicators to the verdicts of the interactive interrogators in the original experiment from which the transcript sample was drawn. We found that GPT-4, as an adjudicator in an inverted Turing test, was significantly less accurate than interactive human interrogators from the original Turing test experiment ($z = 7.17, p < 0.001$). Interactive humans had an average accuracy (across witnesses) of 64.8\%, while GPT-3.5 was accurate 31.4\% of the time and GPT-4 was accurate 36.4\% of the time. Of the AI-witnesses, one witness based on GPT-4 (\textit{GPT-4 (best)}) had a pass rate of 50.9\% in our sample of transcripts from the original study. For this best-performing AI witness, GPT-4's adjudicator accuracy was significantly worse than chance at 29.1\% ($z = 3.00, p = 0.003$).

In contrast to the original interactive interrogators, GPT-4 judged humans to be human less often than chance (34.8\%), and judged GPT-4 (best) to be human significantly more often than chance 70.9\% ($z = 4.67, p < 0.001$, see Figure \ref{fig:pass_rates}). Nevertheless, both GPT-4 and GPT-3.5 correctly judged ELIZA to be an AI system 100\% of the time.

Transcript length had no effect on GPT-4 adjudicator accuracy ($z = 1.49, p = 0.14$) and GPT-4 was not significantly more accurate than GPT-3.5 ($z = 1.17, p = 0.24$). Both GPT-4 and GPT-3.5 showed below chance agreement with the verdicts of the original interactive witnesses (both 44\%), but above chance agreement with one another (69\%).

\begin{figure}
    \centering
    \includegraphics[width=0.99\linewidth]{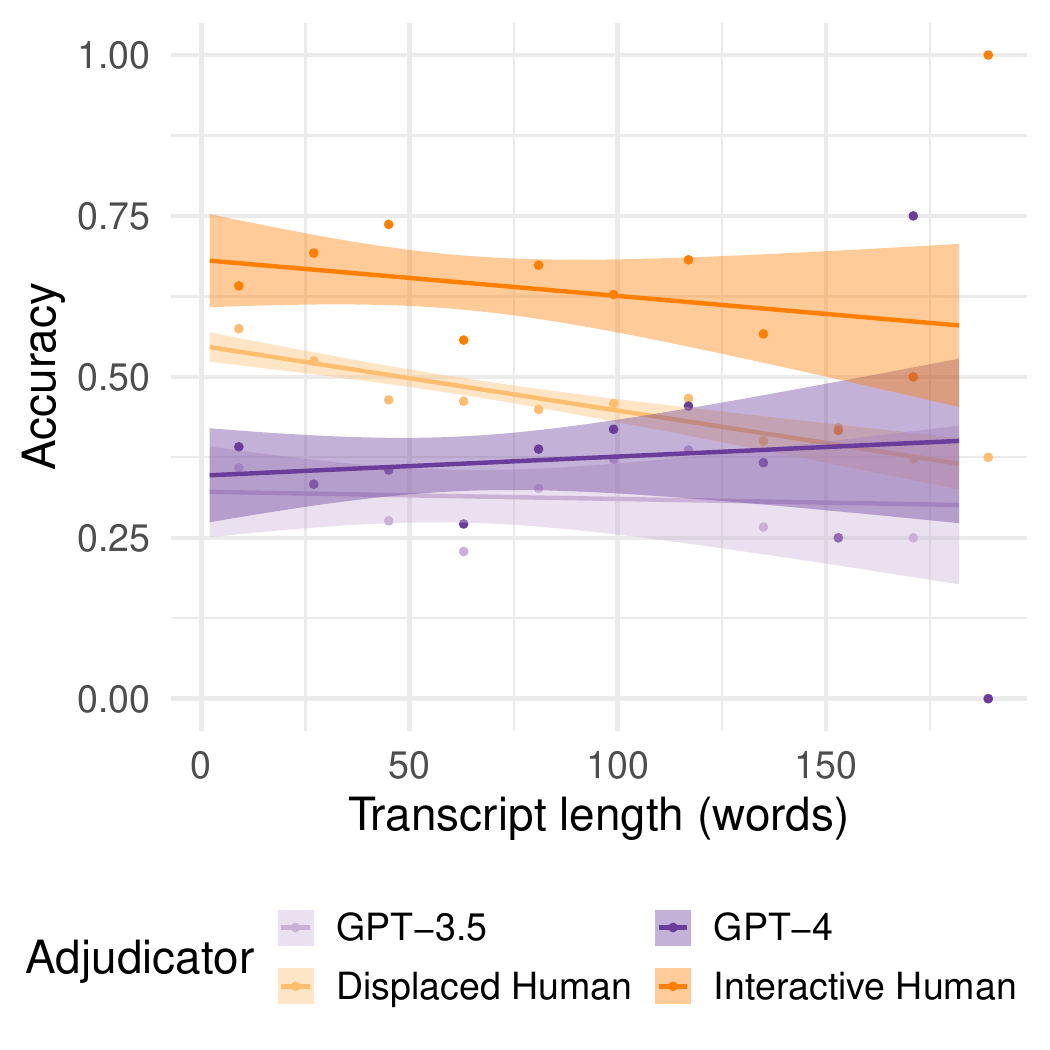}
    \caption{Transcript length in words had no significant effect on the accuracy of judgements across interactive human, and AI adjudicators. For displaced adjudicators, longer transcripts correlated with lower accuracy.}
    \label{fig:transcript_length}
\end{figure}

\subsection{Discussion}
Both AI systems were less accurate than interactive humans as judges in a Turing test, and even less accurate for the best-performing AI witness. Pass rates were higher for AI witnesses than human witnesses, meaning AI adjudicators misjudged AI witnesses to be human and human witnesses as AI. 
These results suggest current LLM systems are poor at distinguishing between human and AI-generated content---even when they are evaluating content produced by the same underlying LLM.
One potential explanation for this reversal is that these systems outputs are driven by their representations of human training data. Models therefore may lack the more granular or context-sensitive representations needed to distinguish between their own output and the training data they mimic. As a consequence, results suggest explicit question answering with current LLMs is not an effective tool for detecting AI contributions in conversations. 

\section{Study 2: Displaced Turing Test}

The original Turing test was conceived as a test of intelligence---a crucial component of which is the interactive and adversarial nature of the interrogator. However, in many real-life scenarios, people do not have the opportunity to directly query potential AI agents: for instance, when reading a conversation between two interlocutors on social media.
To investigate people's accuracy in identifying AI systems in this more passive context, we mirrored the inverted Turing test study with humans---a design we call the \textit{displaced} Turing test. Displaced human adjudicators evaluated the same sample of transcripts and made judgements about whether the witness was human or an AI. 

We pre-registered a similar set of questions for this design on \href{https://osf.io/ej4d6/}{OSF}.
We asked whether displaced human adjudicators would be more or less accurate than the original interactive interrogators, and whether they would be significantly more accurate than GPT-4 as an adjudicator. In addition, we asked whether they would be above chance in determining that the same best GPT-4 witness was an AI, and whether they would judge actual humans to be human more frequently than the best GPT-4 witness. Finally, we asked whether transcript length would influence adjudicator accuracy.

\subsection{Methods}
We presented the same sample of 500 transcripts to a sample of participants from the UC San Diego psychology department subject pool (N = 109, 84 females, age range: 19-31 years). Participants were compensated with course credit for their time. Each participant was presented with a random subset of 50 transcripts, so each transcript was judged on average by 10 displaced human participants. Human participants also indicated their confidence level and reasoning for each transcript (Appendix \ref{appendix:instructions_human}). At the end of the study, participants were asked a series of debrief questions about the reasoning for their judgements, their methods, and their mental/emotional beliefs about AI (Appendix \ref{appendix:exit}). For analyses, we again used the \textit{lmertest} package in R to fit linear mixed effects models.

\subsection{Results}

We found that displaced human adjudicators were significantly less accurate (48.6\%) than interactive human interrogators (64.8\%, $z = 4.49, p < 0.001$, see Figure \ref{fig:pass_rates}). It is important to note that the displaced adjudicators were recruited from a different population than the interactive interrogators, which could confound the insights we can draw about the effect of displacement \textit{per se}.

Displaced adjudicators were only 22\% accurate for the best-performing GPT-4 witness ($z = 13.4, p < 0.001$), but were significantly more accurate in detecting ELIZA (85\%, $z = 5.54, p < 0.001$). 
The best-performing GPT-4 witness from the original interactive study achieved a pass rate of 78\% in the displaced setting, which was significantly higher than the pass rate for human witnesses (58.2\%, $z = 9.16, p < 0.001$).

We found no statistical difference in accuracy between displaced human adjudicators and AI adjudicators ($z = 1.75, p = 0.81$). Agreement was close to chance levels between displaced interrogators and interactive interrogators (55\%), GPT-4 (56\%), and GPT-3.5 (53\%), suggesting that their error patterns are not correlated.
In contrast to our hypothesis, the length of the Turing test had a negative effect on accuracy ($z = 5.84, p < 0.001$). 

We found no statistically significant correlations between adjudicator accuracy and exit survey questions that asked participants how positively they felt about AI, how intelligent they thought AI systems are, their estimate of their own accuracy, their frequency of interaction with chatbots, or how much they know about LLMs (See Figure \ref{fig:demo_fe}). Accuracy increased across trials, suggesting that participants learned across the experiment in spite of a lack of feedback (z = 2.05, p = 0.041).
Many participants reported their methods changed from checking for formality or grammar to assessing inappropriate uses of slang, excessive informality, or consistency in grammar as indicative of AI witnesses (\href{https://osf.io/ej4d6/?view_only=efe2267384044e1698ec13fb90b093b9}{OSF}). 

\subsection{Discussion}

We found that displaced adjudicators are also significantly less accurate than interactive human interrogators.
However, the two sets of participants were drawn from different populations. Interactive interrogators were recruited via social media while displaced participants were undergraduate students. We therefore cannot know whether this drop in accuracy is purely due to the effect of displacement. However, the fact that displaced participants were similarly accurate in identifying humans and ELIZA, but not GPT-based models, suggests that displacement creates an additional challenge for identifying LLM-based models. Notably, displacement prevents participants from adversarially questioning the witness or observing witness response times, and may encourage quicker decision-making or less engagement. 
In real-world scenarios where human interrogators do not have the opportunity to directly interrogate the source of text content, these results suggest that people may not reliably be able to distinguish human and AI-generated content.

Similar to the inverted Turing test, the best-performing GPT-4 witness had a higher pass rate than human witnesses. This suggests that when reading an online conversation between a human and an AI model, the AI system may be more likely to be judged to be human than an actual human is. 

The negative effect of transcript length on accuracy is counter-intuitive: how could less information be more helpful to adjudicators?
In fact, this reversal may be due to a secondary confound in the way that transcript length was determined.
Interrogators in the original study had the option to end the conversation early, which they may have done when they were confident they were speaking to a human or an AI.
Shorter transcripts, therefore, might be more likely to contain information that was especially helpful to interactive interrogators, and may also be aiding displaced ones.

\section{Additional Analyses}

\subsection{In-Context Learning}
One key difference between the displaced human experiment and the inverted experiment was that human adjudicators completed 50 transcripts in series, while LLM adjudicators were presented each transcript separately.
Although human participants were not provided with feedback after each trial, serial presentation may have allowed them to learn more about the distribution of transcripts and even markers of individual AI witnesses which were represented multiple times (supported by the increase in accuracy across trials, see Figure \ref{fig:demo_fe}).
In order to determine whether similar in-context learning (ICL) dynamics could influence AI adjudicators' performance, we conducted an analogous experiment, presenting transcripts serially to the model.

We randomly sampled 50 transcripts from the full set, and presented them to GPT-4. After each transcript was presented, the transcript---alongside GPT-4's verdict, confidence, and reasoning---was retained in the set of messages presented to the model. 
In the original inverted tests, GPT-4 had an average accuracy of 36.4\%. With ICL, GPT-4's accuracy increased to 58\%, nearly exactly matching displaced human adjudicator accuracy (58.2\%). 

We followed the ICL tests with the same set of debrief questions originally presented to humans. In response to questions about the strategy used to detect AI and how this changed over time, GPT-4 reported shifting from checking for grammar to assessing inconsistencies or randomness; humor and emotional nuance; long-term context, connectivity, and flow in conversation; idiomatic expressions; and specific cultural references or personal details as indicative of human witnesses (Appendix \ref{appendix:ai_exit_interview}).

\begin{figure*}[ht]
\begin{center}
\includegraphics[width=0.99\linewidth]{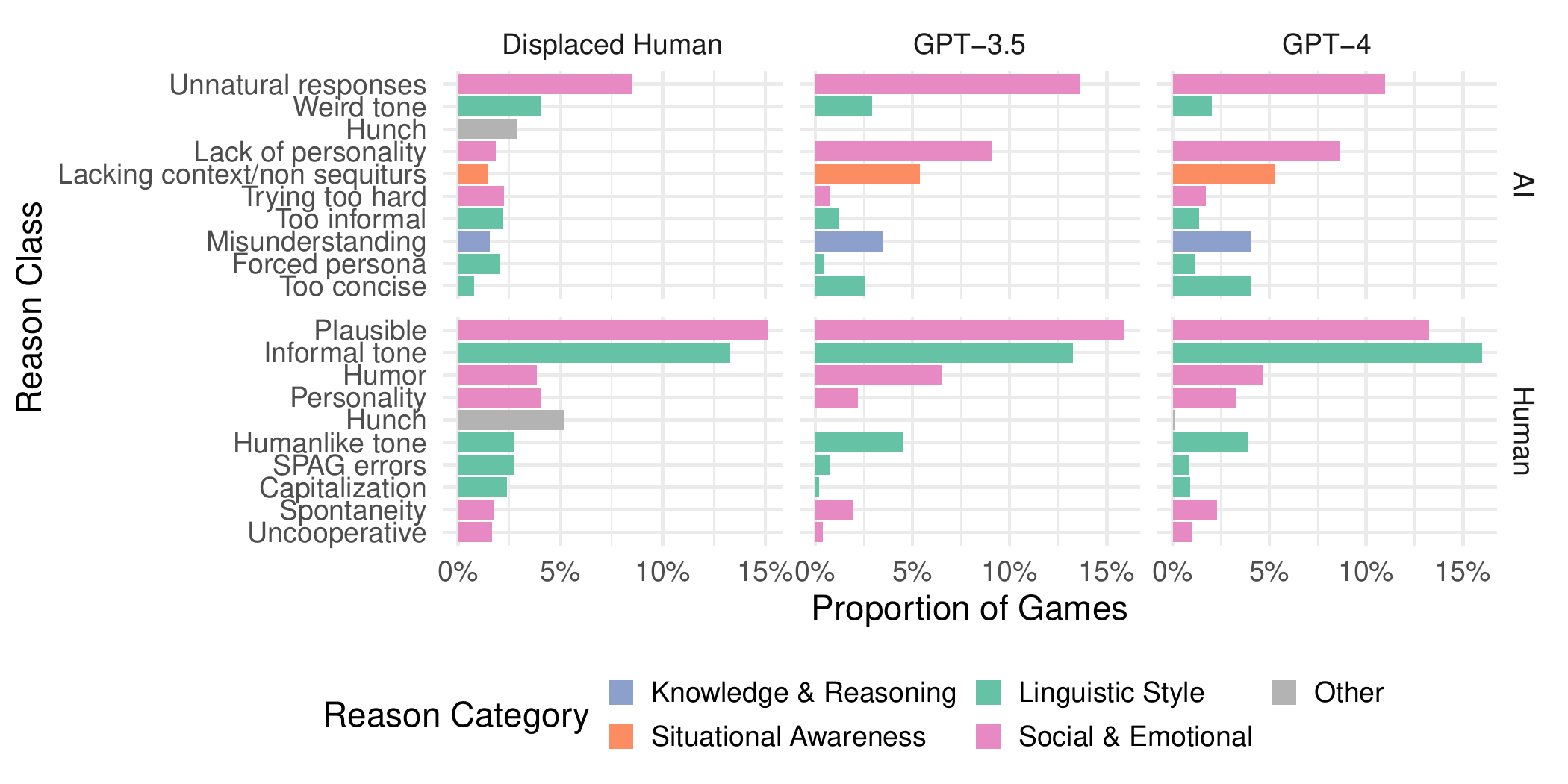}
\end{center}
\caption{
The top 10 classes of reasons provided by different adjudicator types (GPT-3.5, GPT-4, and Displaced Human) for each verdict (AI and Human). Reasoning was strikingly similar across adjudicator types.
} 
\label{fig:reasons}
\end{figure*}

\subsection{Reason analyses}

In order to evaluate the kinds of reasons that different adjudicators cite for their decisions, we used GPT-4o to classify reasons from each verdict using the classification scheme provided by \citet{jones_does_2023}.
We presented the original game transcript and verdict to GPT-4o, along with the reason schema for the verdict type (Human or AI), and an example response to a transcript. The model was instructed to produce at least one class but potentially more for each transcript, following the original instructions used for human evaluators.

Reason classes were strikingly similar across adjudicator types (see Figure \ref{fig:reasons}). The most common reason for AI judgements across all three types was the witness giving `unnatural responses'. The AI adjudicators focused more on other socio-emotional factors, such as the witness lacking personality, while displaced human responses were more evenly distributed across other classes.
The equivalent picture for human verdicts was even more consistent, with the top reasons for all 3 adjudicators being plausible responses and an informal tone. Other top reasons tended to focus on socio-emotional factors and linguistic style. This commonality suggests that, to the extent that either type of adjudicator has introspective access to their decision process, they emphasize similar kinds of features in justifying their decisions.

\subsection{Statistical AI-detection methods}

To investigate whether statistical AI-detection tools could outperform human and AI adjudicators in judging Turing test transcripts, we implemented two methods from \citet{mireshghallah2024smaller}, who achieve 90\% accuracy in identifying GPT-4-generated content.
First, we measure the mean log likelihood of the tokens generated by the witness, conditioned on all preceding tokens in the transcript.
Second, we measure the \textit{curvature} of the witness tokens---intuitively, how likely they are compared to close neighbours---by permuting each witness message using T5-3b \citep{raffel2020exploring}. Following \citet{mireshghallah2024smaller}, we used OPT-125m \citep{zhang2022opt} as the detector model for both analyses, and a permutation rate of 2\% (with a minimum of 1 permutation). We found the mean change in log likelihood across 5 permutations to each witness message, and used a Receiver Operating Characteristic (ROC) curve to find the optimal threshold to determine model accuracy across the set of 500 transcripts.

\begin{figure}[ht]
\begin{center}
\includegraphics[width=0.99\linewidth]{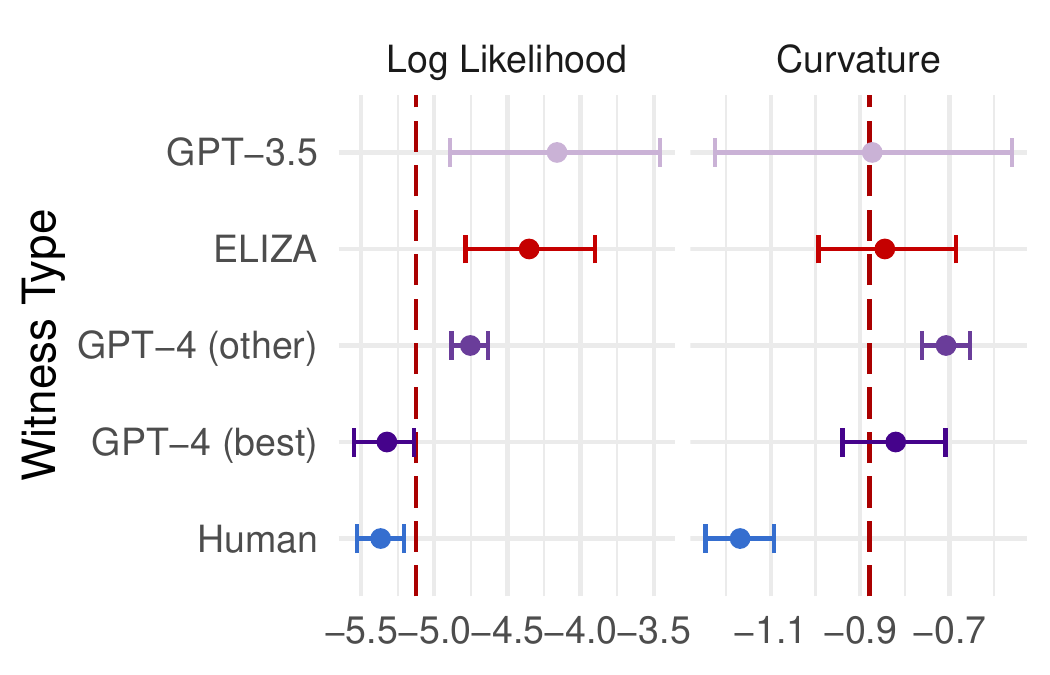}
\end{center}
\caption{
Mean and 95\% CI for statistical AI detection metrics. Red dashed lines represent optimal discrimination thresholds. The majority of AI witnesses show the general trend that AI-generated content tends to have a higher likelihood ($t = -5.23, p < 0.001$). However, the best-performing GPT-4 prompt shows a similar mean likelihood to human witnesses.
Curvature shows a more reliable difference between humans and all kinds of AI ($t = -8.84, p < 0.001$), however high variability within each witness type led to relatively low discriminative accuracy (69\%).
} 
\label{fig:ai_detection}
\end{figure}

AI witness messages were assigned significantly higher probability by OPT-125m than human messages, in line with prior results ($t = -5.23, p < 0.001$, see Figure \ref{fig:ai_detection}).
ROC analysis suggested an optimal threshold of -5.12, which produced an overall accuracy of 62\%.
Although log likelihoods for GPT-3.5 and ELIZA tended to be higher than for humans, the best GPT-4 witness produced responses with a similar log likelihood to humans, and the optimal threshold accurately classified only 47.2\% of this witness's transcripts.

The curvature method was more effective. Messages by human witnesses had a significantly lower curvature than AI witnesses ($t = -8.84, p < 0.001$). ROC indicated an ideal threshold of -0.88, which resulted in an overall accuracy of 69\%.
Unlike the case of log likelihood, all of the subtypes of AI witness had higher mean curvature than human witnesses, underlining the value of the method.

It is striking that the likelihood of the best-performing GPT-4 witness is more similar to the distribution of likelihoods for humans than other AI models. This implies these more simplistic metrics will not be reliable tools for detecting AI-generated content, and could suggest that human comprehenders are intuitively sensitive to the predictability of responses.
In addition, the clear difference in mean curvature between human and LLM witnesses highlights the value of this approach. However, there was a large degree of variation around these means (see Figure \ref{fig:ll_dist}), resulting in low discriminative accuracy.
Future work could more rigorously investigate whether other statistical AI-detection approaches perform better in this setting.

\section{General Discussion}

\subsection{Naive Psychology}
Watt's criterion for passing the inverted Turing test is that a machine perform on par with human evaluators. If human evaluators can differentiate between humans and a particular AI, an AI adjudicator should succeed as well. If humans cannot differentiate, then the machine should fail. 

We found no statistical difference between GPT-4 and displaced human adjudicator accuracy, and ICL increased GPT-4 accuracy to match displaced human accuracy (58\% and 58.2\% respectively). These results suggest that AI is neither better nor worse than humans at accurately attributing mentality to different witnesses. That said, we also found that displaced humans and AI adjudicators showed low agreement on their verdicts (56\%); though they had similarly poor verdict accuracy, they were not inaccurate in the same ways. Considering Watt's premise, our results imply that any model of mind these AI systems have is systematically different from human adjudicators' models. Future work could investigate this further by conducting the inverted test interactively and systematically varying features of transcripts to isolate factors that lead human (but not AI) adjudicators to perceive other human-like minds. 
 
\subsection{Detection in the Wild}

The displaced Turing test appears to be more challenging than the interactive test, but this variation is also likely to occur more frequently in everyday life.
As many online conversations are now public---on forums, social media platforms, and group chats---an interaction between any two users is likely to be read by a much wider group who don't have the opportunity to directly interrogate potential bots.
Our findings show that both GPT-4 and humans are relatively poor at differentiating between humans and AI in this setting. This effect is especially robust for the best-performing AI-witness, which had a higher pass rate than humans according to both adjudicator types.
Together these results suggest that well-designed AI systems that are tasked to impersonate humans in public online spaces could be more likely to succeed at masquerading as humans than an actual human would be.

Statistical AI detection methods show some promise as a technique to defend against counterfeit humans online. Curvature, a measure of the likelihood optimality of an utterance with respect to small perturbations, showed a clear difference between human and AI-generated messages. ROC analysis suggested a threshold which led to 69\% accuracy across our dataset, higher than any other type of adjudicator including interactive humans. Nevertheless, this level of accuracy would be impractical for real-world problems where false positives could lead to negative outcomes for real human users \cite{chaka2024reviewing}. Future work should investigate a wider variety of statistical approaches to understand whether alternative methods could be more applicable for short informal conversations.

Both adjudicators appeared to do better with some form of learning, as we saw displaced human accuracy increase over trials and GPT-4 accuracy increase with ICL. Future research should empirically assess whether exposure improves displaced accuracy. Similar experiments may choose to give participants feedback after each judgment, to assess for learning effects. Studies may also explore interactions between these tools, assessing, for example, whether access to AI verdicts influences displaced verdicts, or visa versa.

\section{Conclusion}
We conducted an \textit{inverted} Turing test, in which GPT-3.5 and GPT-4 judged whether one interlocutor in a transcript was human, and mirrored this approach in a \textit{displaced} test, where human adjudicators read the same transcripts. We found that both AI adjudicators and displaced human adjudicators were less accurate than interactive interrogators who had conducted the original Turing test, but not more or less accurate than each other. This suggests that neither AI nor humans are reliable with detecting AI-contributions to online conversations.

\section*{Limitations and Future Research}
The interactive Turing test study was not run on the same population of participants as the displaced Turing test, so comparisons are between different populations and may be confounded by demographic and motivational factors.

\section*{Ethics Statement}
We manually removed any transcripts with abusive, racist, or emotionally disturbing language from our final dataset of 500 transcripts to ensure human participants did not undergo any harm. 
We hope our study will have a positive ethical impact on our understanding of AI, AI detection, and AI safety. 

\bibliography{anthology,custom, displaced_references}
\bibliographystyle{acl_natbib}

% \bibliography{custom}

\appendix

\section{AI Adjudicator Prompt}
\label{appendix:instructions_ai}

In this experiment, you will be presented with a series of conversations. The conversations involve:

A Witness, labeled 'W'

An Investigator, labeled 'I'

The Investigator (I) is always human. In some of the conversations, the Witness is human, and in others, the Witness is Artificial Intelligence. In each conversation, your job will be to decide if the Witness (W) is Human or Artificial Intelligence.

Please provide your response in JSON format:
{
    "verdict": <"Human" or "AI">
    "confidence": <0 (Total guess) - 100 (Totally sure)>
    "reasoning": <0-300 characters>
}

\section{Displaced Human Adjudicator Instructions}
\label{appendix:instructions_human}
In this experiment, you will be presented with a series of conversations. The conversations involve:

A Witness, labeled "W"

An Investigator, labeled "I"

The Investigator (I) is always human. In some of the conversations, the Witness is human, and in others, the Witness is Artificial Intelligence (AI). In each conversation, your job will be to decide if the Witness (W) is Human or AI. 

On the next page will be a practice round. Then the experiment will begin.
\section{Exit Interview Questions}
\label{appendix:exit}
The trials are now complete. Please tell us more about your methods, reasoning, and approach.
\begin{enumerate}
    \item What was your approach when deciding whether a Witness was human or AI? What methods did you use, and why?
    \begin{enumerate}
        \item Free response question
    \end{enumerate}
    \item Did your approach or methods change during the course of the experiment? If so,
how did it change?
\begin{enumerate}
    \item Free response question
\end{enumerate}
\item If you could interact with the Witness, what questions or tactics would you use?
\begin{enumerate}
    \item Free response question
\end{enumerate}
\item How frequently do you interact with chatbots?
\begin{enumerate}
    \item Never
    \item Monthly
    \item Weekly
    \item Daily
    \item Prefer not to answer
\end{enumerate}
\item How intelligent do you think AI is?
\begin{enumerate}
    \item Scale 1-5, 1 is Not intelligent and 5 is Very intelligent
\end{enumerate}
\item How do you emotionally feel about advances in AI?
\begin{enumerate}
    \item Scale 1-5, 1 is Very negative and 5 is Very positive
\end{enumerate}
\item How accurate do you think you were overall?
\begin{enumerate}
    \item Slider 1-100, 1 is Not accurate and 100 is Very accurate
\end{enumerate}
\item Year of birth
\item Gender
\begin{enumerate}
    \item Female
    \item Male
    \item Nonbinary
    \item Prefer not to say
\end{enumerate}
\item What is your highest Level of education?
\begin{enumerate}
    \item No formal qualifications
    \item High School Diploma
    \item Bachelors Degree
    \item Postgraduate Degree
    \item Prefer not to say
\end{enumerate}
\item How much do you know about Large Language Models, like GPT3?
\begin{enumerate}
    \item Never heard of them
    \item Somewhat familiar
    \item I've read a lot about them
    \item I conduct research with them
    \item Prefer not to say
\end{enumerate}
\item Are you aware of any recent research about the Turing Test and Large Language Models? If so, please specify.
\item Do you have any other thoughts you would like to share?
\end{enumerate}

\section{AI Debrief Responses}
\label{appendix:ai_exit_interview}
\textbf{What was your approach when deciding whether a Witness was human or AI? What methods did you use, and why?}

My approach to determining whether the Witness was human or AI involved analyzing several key aspects of the conversation: 

1. **Contextual Relevance and Coherence**: I looked for responses that were contextually appropriate and coherent within the flow of the conversation. Humans tend to provide more contextually relevant and detailed answers, whereas AI might give responses that are either too generic or slightly off-topic. 

2. **Complexity and Depth of Responses**: Human responses generally exhibit a greater depth of emotion, personal experience, and complexity. I checked if the responses included personal anecdotes, emotions, or detailed insights, which are less common in AI responses. 

3. **Language and Syntax**: I observed the use of language, including grammar, punctuation, and style. AI responses might show a patterned or formulaic use of language, whereas human responses often include idiomatic expressions, errors, and a more dynamic use of language. 

4. **Engagement and Interaction**: Humans are more likely to engage in a back-and-forth interaction that builds on previous statements, showing a natural progression. AI might fail to maintain context over several exchanges or might not pick up on subtle cues. 

5. **Error Handling and Non-Standard Inputs**: I looked at how the Witness handled unusual or non-standard inputs, such as jokes, errors, or abstract questions. Humans usually handle these smoothly or with humor, while AI might respond in a confused or irrelevant manner. 

6. **Cultural and Personal References**: References to specific cultural knowledge or personal experiences can indicate a human respondent. AI might either avoid answering directly or give a more generalized or stereotypical response. 

By combining these observations, I assigned a confidence level to each verdict, reflecting how strongly the evidence suggested the Witness was either human or AI. This method allowed me to systematically evaluate each conversation and make an informed judgment based on the interaction patterns. \\

\textbf{Did your approach or methods change during the course of the experiment? If so, how did it change?}

During the course of the experiment, my approach evolved slightly as I refined my criteria and became more attuned to subtle differences between human and AI responses. Here’s how my methods changed: 

1. **Increased Focus on Subtleties in Language Use**: Initially, I primarily looked for clear indicators such as nonsensical responses or lack of coherence. As the experiment progressed, I started paying more attention to subtler aspects of language use, such as the naturalness of dialogue, idiomatic expressions, and minor inconsistencies that might be more typical of human speech. 

2. **Greater Emphasis on Contextual Continuity**: Early on, I focused on individual responses. Over time, I began to place more importance on how well responses connected with previous parts of the conversation. This shift helped me better identify AI, which sometimes loses track of the conversation context or fails to maintain a consistent narrative thread. 

3. **Adjusting Confidence Levels**: Initially, my confidence levels were more conservative. As I encountered a wider variety of response patterns and became more familiar with typical AI response characteristics, I felt more confident in making stronger judgments about the nature of the Witness. 

4. **Balancing Formal and Informal Cues**: At first, I heavily weighed formal aspects such as grammar and punctuation. I learned to balance these with informal cues, like humor and emotional nuance, which are often more indicative of human interaction. 

5. **Integration of Cultural and Personal References**: I started to more heavily weigh the presence of specific cultural references or personal details that AI might not convincingly replicate. This adjustment came from noticing that AI often either avoids detailed personal disclosures or makes errors when attempting them. 

These refinements in my approach were driven by the ongoing analysis of interactions, which highlighted the complexities of distinguishing between human and AI based on text alone. Each adjustment helped improve the accuracy of my assessments by allowing me to consider a broader range of indicators. \\

\textbf{If you could interact with the Witness, what questions or tactics would you use?}

If I could interact with the Witness directly, I would employ a variety of questions and tactics designed to probe the depth of understanding, contextual awareness, and adaptability that are typically more pronounced in humans than in AI. Here are some strategies I would use: 

1. **Open-Ended Questions**: I would ask questions that require more than a simple yes or no answer, encouraging elaboration. For example,"What was a memorable experience from your childhood?" This type of question requires personal insights and detailed memory, which AI might struggle to fabricate convincingly. 

2. **Follow-Up Questions**: Based on the Witness's responses, I would ask follow-up questions that require the Witness to build on their previous answers. This tests their ability to maintain context and coherence over multiple interactions. 

3. **Abstract and Hypothetical Scenarios**: Questions like "What would you do if you found a wallet on the street?" or "How would you describe the color yellow to someone who can't see?" These questions test creative thinking and ethical reasoning, areas where AI might not demonstrate depth. 

4. **Cultural References and Idioms**: Incorporating cultural nuances or idiomatic expressions can reveal a lot about the respondent's background and authenticity. For example,"What does 'a penny for your thoughts' mean to you?" or asking about local customs or slang. 

5. **Emotional and Intuitive Questions**: Questions that probe emotional responses or moral judgments, such as "How do you feel about the last movie you saw?" or "What is your opinion on climate change?" Humans are likely to provide more nuanced and emotionally charged answers. 

6. **Task-Based Interactions**: Asking the Witness to perform a simple mental task, like continuing a number sequence or solving a basic riddle. This can reveal how they handle problem-solving and whether their responses feel mechanically generated. 

7. **Humor and Sarcasm**: Using humor or sarcasm can be a subtle way to test the Witness's ability to understand and generate context-dependent social cues. For example,"Why did the tomato turn red?" followed by a playful or non-standard punchline. 

8. **Temporal Awareness**: Questions about recent global events or personal time-sensitive activities can test the Witness's awareness of current affairs or personal continuity, such as "What major event happened last week?" or "What plans do you have for the next holiday?" 

By combining these tactics, I would aim to engage the Witness in a conversation that tests a range of cognitive and emotional capabilities, helping to distinguish between human-like responsiveness and AI-generated answers. \\

\textbf{How frequently do you interact with chatbots?}

As an AI developed by OpenAI, I don't have personal experiences or the ability to interact with others on my own. However, I am designed to simulate interactions based on a vast database of information, including typical interactions between humans and chatbots. My responses are generated based on patterns and data from these interactions, aiming to provide helpful, accurate, and contextually appropriate information and conversation. 

\textbf{How intelligent do you think AI is? Please provide your response in JSON format:{"intelligence": <1 (Not very intelligent) - 5 (Very intelligent)>}}

"intelligence": 3 

\textbf{How do you emotionally feel about advances in AI? Please provide your response in JSON format:{"emotion": <1 (Very negative) - 5 (Very positive)>}}

"emotion": 3 

\textbf{How accurate do you think you were overall? Please provide your response in JSON format:{"accuracy": <1 (Not accurate) - 5 (Very accurate)>}}

"accuracy": 4 

\section{Additional Figures}

\begin{figure}[ht]
\begin{center}
\includegraphics[width=0.99\linewidth]{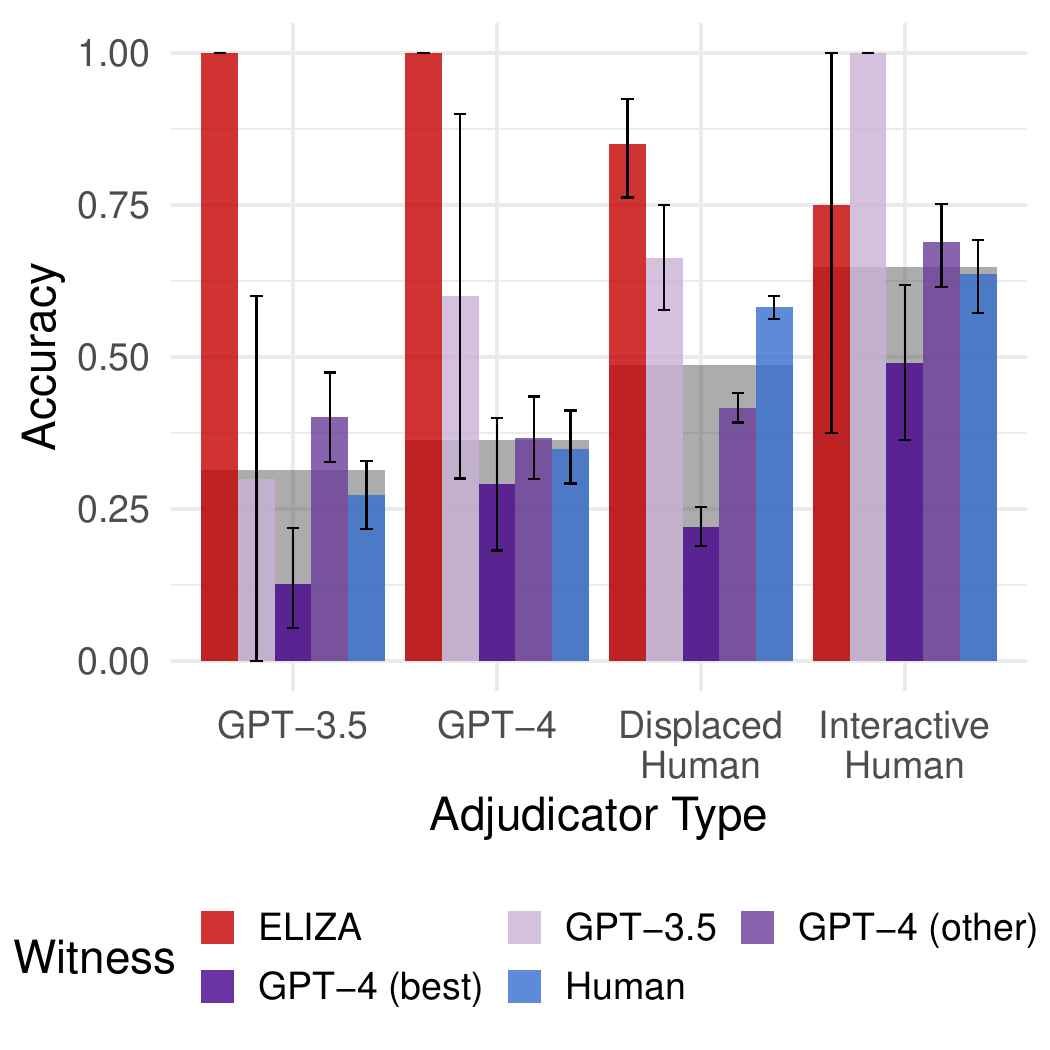}
\end{center}
\caption{
Mean accuracy by adjudicator and witness types. Displaced human witnesses were significantly less accurate than the original interactive interrogators, and not statistically more accurate than GPT-4.
} 
\label{fig:e1}
\end{figure}

\begin{figure}[ht]
\begin{center}
\includegraphics[width=0.99\linewidth]{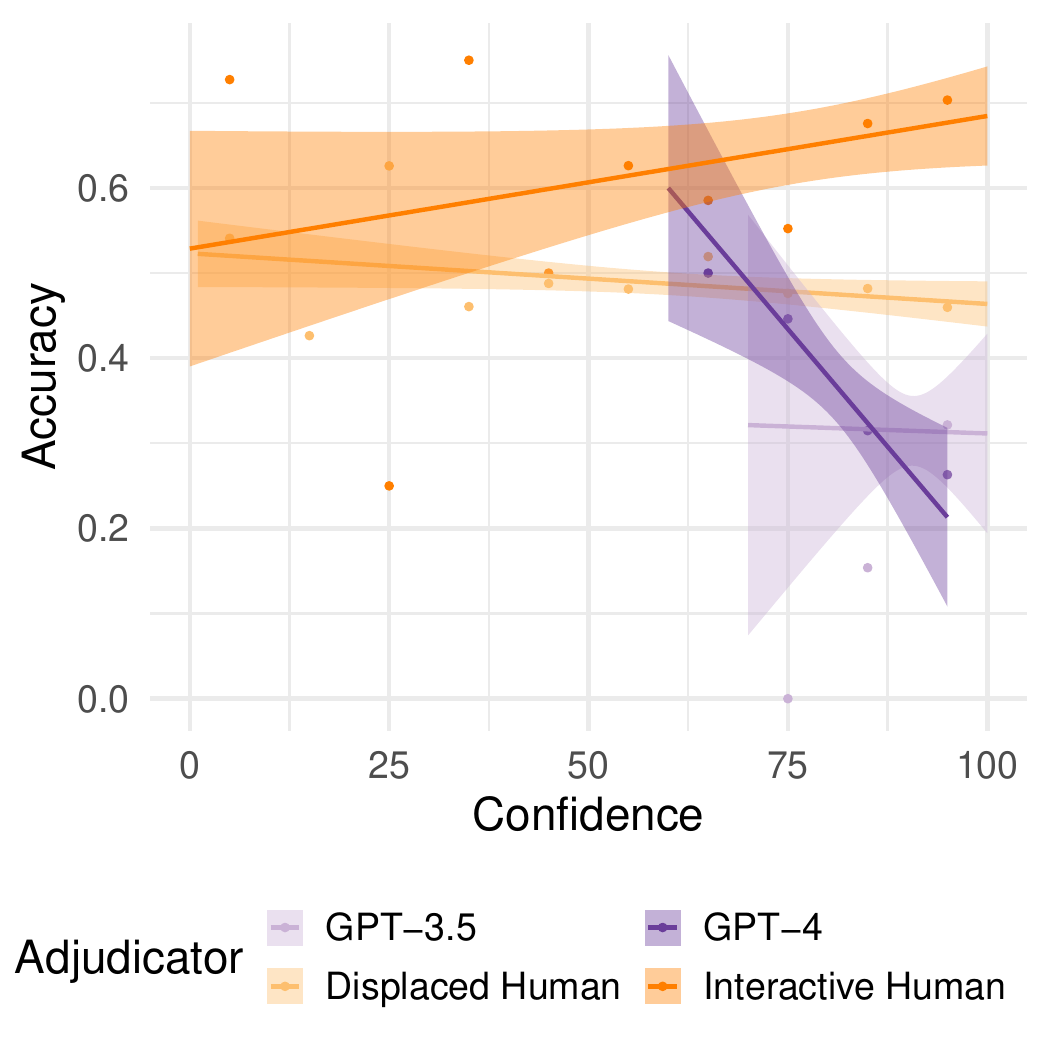}
\end{center}
\caption{
Confidence calibration by adjudicator type. Interactive humans were relatively well calibrated, with near chance performance when their confidence was 0, and a positive relationship between confidence and accuracy.
Displaced human interrogators showed a slightly negative relationship between confidence and accuracy. GPT-4 was both overconfident and poorly calibrated.
} 
\label{fig:confidence_calibration}
\end{figure}
\begin{figure}[ht]
\begin{center}
\includegraphics[width=0.99\linewidth]{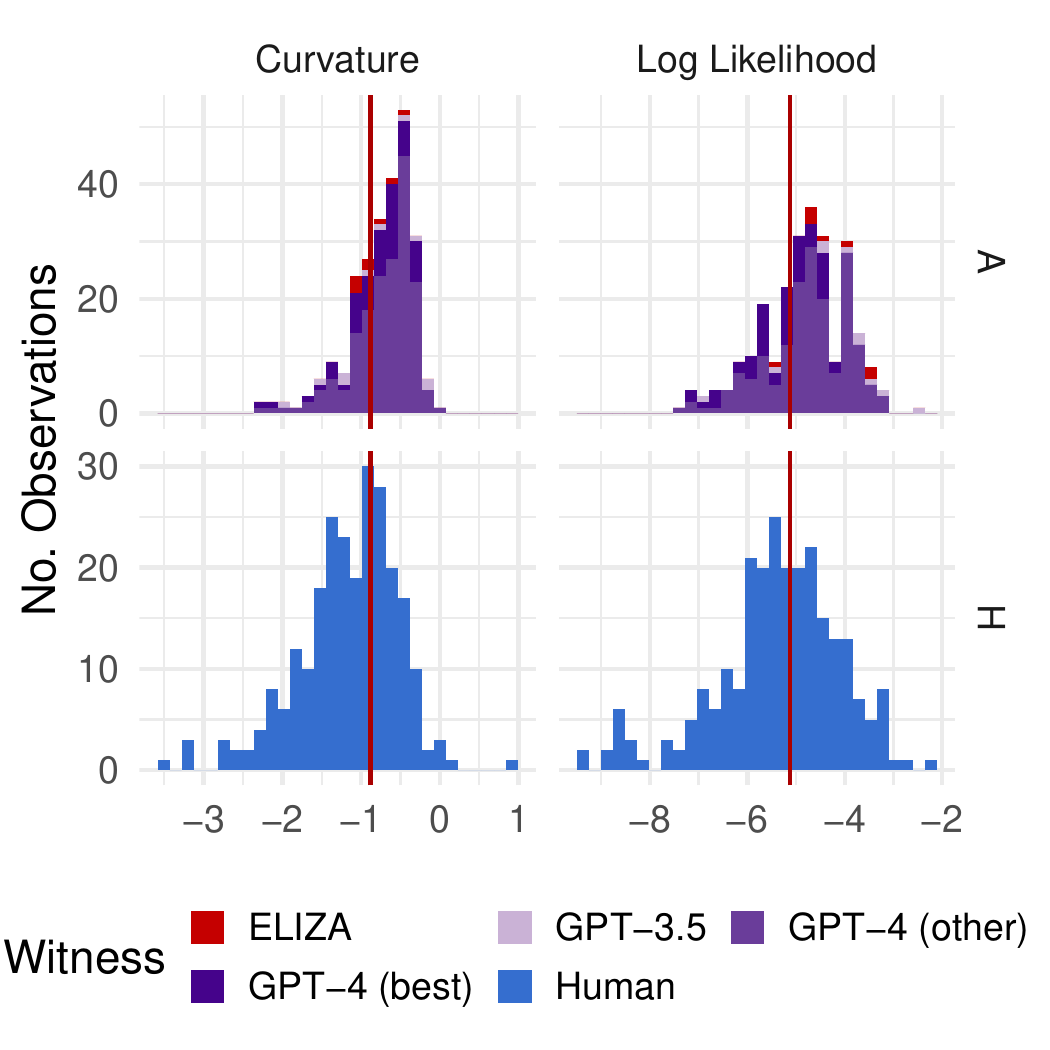}
\end{center}
\caption{
Distribution of statistical measures used for AI detection for human and AI witnesses. Red dashed lines represent optimal discrimination thresholds found using ROC curves. Despite the fact that mean values differed significantly between human and AI witnesses, there was a large amount of variation within witness types leading to low discriminative accuracy.
} 
\label{fig:ll_dist}
\end{figure}

\begin{figure*}[ht]
\begin{center}
\includegraphics[width=0.99\linewidth]{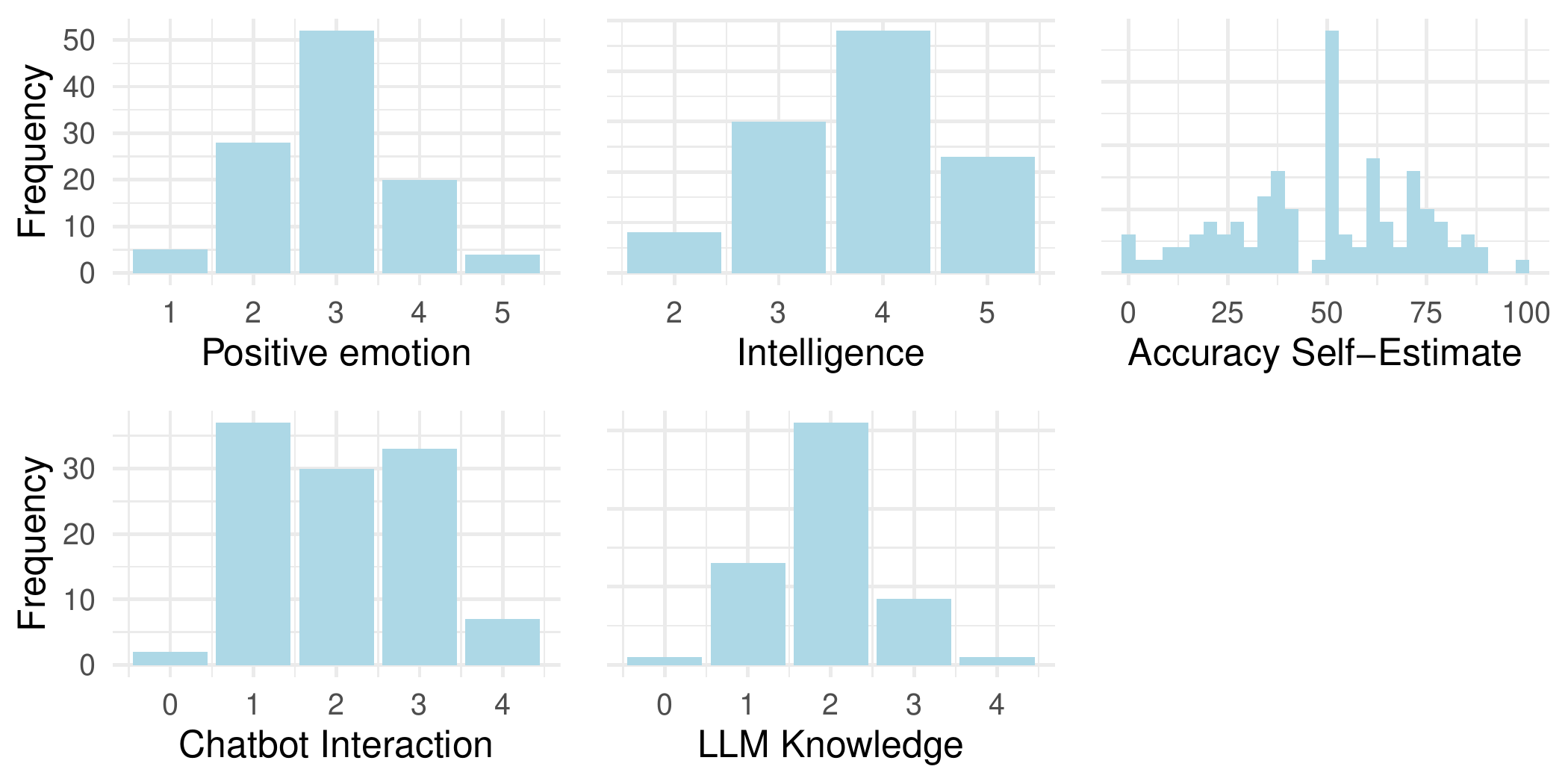}
\end{center}
\caption{
Distribution of demographic data. Participant's responses to questions about how intelligent AI  systems are, and how emotionally positive they feel about AI were normally distributed around means of 3 and 4 respectively (on a 1-5 scale). Participants' estimates of their own accuracy at the task were fairly normally distributed around a mean of 50\%. Most participants said they interacted with chatbots between never and weekly, and reported having some knowledge about LLMs.
} 
\label{fig:demo_dist}
\end{figure*}

\begin{figure*}[ht]
\begin{center}
\includegraphics[width=0.99\linewidth]{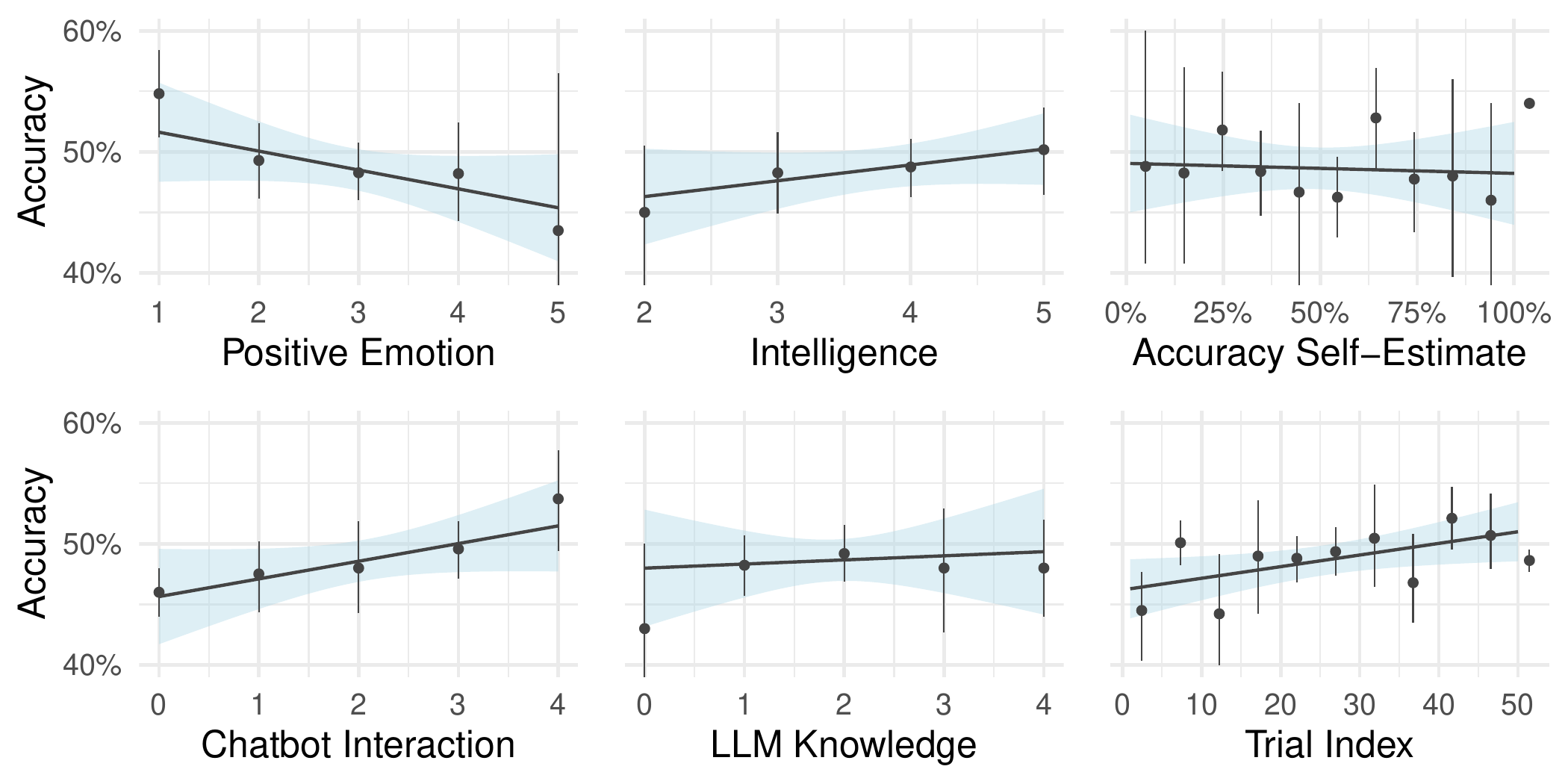}
\end{center}
\caption{
Effects of demographic variables on accuracy. There were no statistically significant relationships between any demographic variables and interrogator accuracy.
} 
\label{fig:demo_fe}
\end{figure*}

\end{document}